# Mechanistic Insights into Water-Splitting, Proton Migration, and Hydrogen Evolution Reaction in g-C$_3$N$_4$/TiO$_2$-B and Li-F co-doped Heterostructures

Shuhan Tang,[a] Qi Jiang,[a] Shuang Qiu,[a] Hanyang Ji[a,b] and Xiaojie Liu[a,b,*]

Solar water splitting has received a lot of attention due to its high efficiency and clean energy production potential. Herein, based on the band alignment principle, the g-C$_3$N$_4$/TiO$_2$-B(001) heterostructure is strategically designed, then a Li-F co-doping approach is developed and implemented, leading to significant enhancement in the photocatalytic hydrogen evolution efficiecy of the heterostructure systems. The decomposition of water molecule on the surface of heterostructures, the migration and diffusion of proton across the interface, and the hydrogen evolution performance are systematically studied and comprehensively analyzed. The results demonstrate that the heterojunction surface exhibits a relatively low energy barrier for water decomposition, facilitating both hydrogen evolution reaction (HER) and oxygen evolution reaction (OER). Proton transfer preferentially occurs from the TiO$_2$-B(001) surface to the g-C$_3$N$_4$ surface through the interface. The presence of polar covalent bonds establishes a substantial energy barrier for proton migration from TiO$_2$-B(001) surface to the interface, representing a rate-determining factor in the hydrogen evolution process. The formation of hydrogen bonds (O-H⋯N) significantly reduces the migration energy barrier for protons crossing the interface to the g-C$_3$N$_4$ surface. Notably, in the Li-F co-doped heterostructure system, while proton diffusion is observed at the interface, this interfacial proton diffusion does not constitute a determining factor for the overall HER performance. Furthermore, Li-F co-doping enhances interfacial polarization, leading to a substantial reduction in the migration energy barrier for proton from interface to the g-C$_3$N$_4$ surface. Hydrogen adsorption free energy analysis show that that the heterojunction surface exhibits optimal proton adsorption and desorption characteristics. The synergistic combination of low water decomposition energy barrier, reduced proton migration energy barriers and exceptional HER performance endows both g-C$_3$N$_4$/TiO$_2$-B(001) heterostructure and Li-F codoped g-C$_3$N$_4$/TiO$_2$-B(001) heterojunction with remarkbale potential as efficient HER photocatalyst.

## Introduction

With the rapid advancement of global economy, issues such as energy crisis, environmental pollution and related challenges have become increasingly severe. The reliance on traditional fossil fuels has led to significant greenhouse gas emissions, while the depletion of the fossil fuels has spurred efforts to explore sustainable and clean energy alternatives. Hydrogen, owing to its high calorific value, pollution-free, and the abundance of its raw materials, is considered as one of the most promising clean energy sources. The core of green hydrogen production lies in the development and application of efficient water-splitting technologies. As early as 50 years ago, scientists demonstrated that water could be split into oxygen and hydrogen molecules by irradiating semiconductor electrodes.[1] However, this process has been limited by low efficiency and high cost, significantly hindering the widespread adoption of solar-driven hydrogen production.[2] In recent years, the use of nanoparticle photocatalyst dispersions has opened new avenues for the development and application for solar-driven hydrogen production.[3] Consequently, solar-driven water-splitting is now considered a highly promising method for generating clean and renewable energy.[4]

The essence of photocatalytic water-splitting for hydrogen production is the photoelectric effect of semiconductor materials. When the energy of incident light exceeds or equals the band gap of the semiconductor, photons are absorbed, exciting electrons from the valence band to the conduction band and generating photoinduced electrons-hole pairs. These electrons and holes migrate to the surface of the semiconductor, where they participate in redox reactions and produce oxygen and hydrogen molecules. However, single-component semiconductors are often limited by their high electron-hole recombination rates, which significantly reduce their photocatalytic hydrogen evolution efficiency. Therefore, a large number of semiconductor heterostructure photocatalysts with well-aligned bandstructures have been designed and synthesized,[5-11] demonstrating exceptional performance in photocatalytic hydrogen evolution reaction (HER) performance. Based on our previous studies,[12] TiO$_2$-B, which possesses unique parallel channels facilitating ions migration along the [010] direction, can be coupled with g-C$_3$N$_4$, a material known for its excellent conductivity and high catalytic activity, to form two-dimensional g-C$_3$N$_4$/TiO$_2$-B(001) heterostructure photocatalyst. This heterostructure photocatalyst exhibits excellent HER performance. Furthermore, it has been found that co-doping with Li and F further enhances the photocatalytic hydrogen evolution efficiency of such heterostructures.[13] The staggered band alignment, efficient separation of photoinduced charge carriers, and the improved visible light absorption collectively indicate that g-C$_3$N$_4$/TiO$_2$-B(001) heterostructures are highly promising candidates for photocatalytic hydrogen evolution catalysts.

However, for HER in neutral and alkaline electrolytes, $H_2O$ serves as the primary proton source. It is well known that HER is a subsequent reaction to the oxygen evolution reaction (OER). A complete HER cycle involves several key stesps: the catalytic water-splitting on the surface of the photoelectrode (e.g., water-splitting on $TiO_2$-B(001) surface), the transport of intermediates (e.g., *H) across the photoelectrode/carrier interface, and efficient desorption of hydrogen on the carrier surface (e.g., g-$C_3N_4$ surface). In HER process, the diffusion and migration of protons at the interface plays a key role in determining the overall reaction kinetics. A significant chanllenge lies in identifying the proton transfer pathway between different components within the heterostructure and understanding how to activate proton transfer effectively. These unresolved issues have lead to a substantial gap in the fundamental understanding of HER mechanism, which in turn severely hinders the design and development of high-performance neutral HER catalysts for neutral environments.

In this study, the mechanisms of water-splitting, the proton transport (including surface/interface diffusion and migration), and the free energy changes during the HER process were systematically investigated to elucidate the proton transport mechanism and HER performance. It was found that water can be stably adsorbed on various $TiO_2$-B(001) surfaces (i.e., $TiO_2$-B(001), $TiO_2$-B(001)/g-$C_3N_4$ and Li-F co-doped $TiO_2$-B(001)/g-$C_3N_4$ heterostructures), with a progressively decreasing dissociation energy. Proton migrates from the $TiO_2$-B(001) surface to the g-$C_3N_4$ surface through the interface, where oxygen acts as both proton donor and acceptor, forming strong polar covalent bonds, leading to a relatively large migration energy barrier. Li-F co-doping enhances interfacial polarization, reducing the migration barrier. During the migration from the interface to g-$C_3N_4$ surface, N acts as a proton acceptor, while interfacial hydrogen bonds (O-H…N) further facilitate proton migration, resulting in a small migration barriers. Notably, in the Li-F co-doped heterostructure, proton transport involves both interfacial migration (0.226 eV) and diffusion (0.701 eV), the latter resulting from the cleavage of the O-H…F hydrogen bond, proton rotation, and the cleavage/recombination of the polar O-H covalent bond. Finally, Gibbs free energy analysis confirmed that the heterostructures are favorable to HER adsorption and desorption. The lower water dissociation energy, reduced migration barrier and lower overpotential of the g-$C_3N_4$/$TiO_2$-B(001) heterostructure and the Li-F co-doped heterostructure enable efficient hydrogen evolution. By systematically studying the proton migration pathways and energy barriers in the pristine and Li-F codoped g-$C_3N_4$/$TiO_2$-B(001) heterostructures, this work aims to reveal the physical origins of the differences in HER activity among different systems and identify potential optimization strategies. These findings not only provide new insights for improving HER efficiency but also offer a viable approach to addressing current environmental pollution and energy crises.

## Calculation methods

Total energy and self-consistent electronic structure calculations were carried out using Vienna Ab initio Simulation Package (VASP) software,[14] based on density functional theory (DFT). The generalized gradient approximation (GGA)[15] with Perdew-Burke-Ernzerhof (PBE)[16] functional was used to treat the exchange-correlation interaction. Spin and self-consistent dipole corrections were applied to all the calculations.[17,18] The projector augmented wave (PAW) method was used to describe electron-ion interactions,[19] with the following valence electrons explicitly considered: $2s^22p^5$ for fluorine, $2s^12p^0$ for lithium, $2s^22p^2$ for carbon, $2s^22p^3$ for nitrogen, $2s^22p^4$ for oxygen and $3d^34s^1$ for titanium. The electronic wave functions were expanded in a plane-wave basis set with a kinetic energy cutoff of 400 eV for all calculations. The Brillouin zone was sampled using the Monkhorst-Pack grid method, with a 2×2×1 k-point mesh emloyed for the geometric optimization of the pristine and Li-F co-doped g-$C_3N_4$/$TiO_2$-B(001) heterostructures. A vacuum of 15 Å was used to eliminate interactions between periodic images in the calculations. All heterostructures were fully relaxed, with energy and force convergence criteria set to $1\times10^{-4}$ eV and 0.01 eV Å$^{-1}$, respectively. To determine the proton diffusion and migration path and energy barrier, the climbing image nudged elastic band (CI-NEB) method was emloyed to identify the minimum pathway from the initial state (IS) to the final state (FS).[20] The energy and force convergence criteria were set to $1\times10^{-7}$ eV and 0.03 eV Å$^{-1}$, respectively, in the CI-NEB calculations. All transition states were verified by imaginary frequency analysis.

## Results and discussions

### Water-splitting on different $TiO_2$-B(001) surfaces

It is well established that $TiO_2$ is an effective catalyst for water-splitting. However, water-splitting represents the initial and critical step in both the hydrogen evolution and oxygen evolution process, significantly influencing the subsequent HER and OER performance In addition, the proton source mainly comes from the water-splitting in neutral or alkaline electrolytes for HER. In order to thoroughly investigate the impact of water-splitting on HER performance, the energetics of water adsorption on various $TiO_2$-B(001) surfaces and the kinetics of water-splitting were systematically studied and analyzed as a primary focus.

Fig. 1(a) illustrates the structural features of the $TiO_2$-B(001) surface, the $TiO_2$-B(001)/g-$C_3N_4$ heterostructure and the Li-F co-doped $TiO_2$-B(001)/g-$C_3N_4$ heterostructure. The $TiO_2$-B(001) surface is composed of chains of $TiO_6$ octahedra, which are interconnected through shared edges, while the chains themselves are linked via shared vertices. All exposed titanium atoms on the surface are unsaturated five-coordinated Ti atoms ($Ti^*_{5c}$), and there are three types of non-equivalent oxygen atoms, including unsaturated bi-coordinated oxygen atoms ($O^*_{2c}$), unsaturated tri-coordinated oxygen atoms ($O^*_{3c}$), and saturated tri-coordinated oxygen atoms ($O_{3c}$).[21] Considering that the O-H bond length in water molecule is 0.96 Å and the H-O-H bond angle is 104.51°, and considering that the O-Ti-O bond angles on the TiO$_2$-B(001) surface range from 99° to 103°, water molecule is expected to preferentially adsorb onto the Ti top site via its O atom, with the two H atoms extending horizontally toward adjacent $O_{2c}^*$ and $O_{3c}^*$ sites. Such a adsorption configuration aligns with the O-Ti-O bond angles on the surface. Subsequently, the adsorbed water molecule dissociates into *OH and *H species. In the TiO$_2$-B(001)/g-C$_3$N$_4$ and Li-F co-doped TiO$_2$-B(001)/g-C$_3$N$_4$ heterostructures, it is thought that the water adsorption behavior should be consistent with that on the TiO$_2$-B(001) surface. However, due to interfacial interactions within the heterostructure and the co-doping of Li and F, the Ti atoms on the heterostructure surface are no longer fully equivalent. Nevertheless, the structure retains an approximate mirror symmetry, enabling the analysis to focus on only half of the adsorption sites.

Based on the above analysis, we systematically constructed and optimized possible adsorption configurations and calculated the adsorption energies of molecular adsorption and dissociation adsorption for water on different TiO$_2$-B(001) surfaces. The adsorption energy is defined as the difference between the total energy of the adsorption system and the sum of the total energy of the non-adsorbed system and the adsorbate. According to this definition, a more negative adsorption energy indicates a stronger interaction between the adsorbate and the TiO$_2$-B(001) surface, as well as a more stable adsorption configuration.

The calculation results demonstrate that the adsorption and dissociation behavior of water is consistent with our predictions. Fig. 1(b) illustrates the most stable adsorption and dissociation configurations of water on different TiO$_2$-B(001) surfaces. For example, on the TiO$_2$(001) surface, the H-O-H bond angle of the adsorbed water molecule is 104.82°, matching the O-Ti-O bond angle (101.56°) of the substrate. Additionally, the H-O bond length of the adsorbed water molecule (0.98 Å) is slightly elongated compared to that of a free water molecule (0.96 Å), confirming the interaction between the surface and the water molecule. The adsorption energy of the adsorbed water system is 0.442 eV lower than that of the dissociated system, indicating that the water dissociation process on the TiO$_2$-B(001) surface is endothermic. Analysis of the water dissociation pathway using the CI-NEB method reveals the absence of a transition state, indicating that the water-splitting on TiO$_2$-B(001) surface is a thermodynamic-driven process rather than a kinetic-controlled process. The water-splitting requires an energy of at least 0.442 eV. The water-splitting energy calculated in the current study is in close agreement with the value reported in the literature (0.42 eV).[22]

Based on the adsorption energy analysis, as one can see from Fig. 1(b), the Ti$_{13}$ site ehxibits the lowest adsorption energy for water on the pristine and Li-F co-doped TiO$_2$-B(001)/g-C$_3$N$_4$ heterostructures. Consequently, the dissociation adsorption energy of water molecules was optimized and calculated using Ti$_{13}$ site as the reference configuration. It is found that the dissociation adsoprtion energy is 0.177eV, with an increase of 0.441eV after dissociation, suggesting that water dissociation on the TiO$_2$-B(001)/g-C$_3$N$_4$ surface is also thermodynamically challenging and requires energy input. Similar to the water-splitting on the TiO$_2$-B(001) surface, no transition state was identified using CI-NEB method, confirming that water-splitting on TiO$_2$-B/g-C$_3$N$_4$ heterostructure surface is an endothermic reaction that is relatively facile, requiring a minimum energy of 0.441 eV. Following Li-F co-doping, water molecule adsorption on the heterostructure becomes more stable, with the adsorption energy at Ti$_{13}$ site reaching -0.471eV, which is 0.207 eV lower than that of the pristine heterostructure. Notably, the dissociation adsorption energy shifts to a negative value of -0.133eV, indicating that water-splitting on the Li-F co-doped heterostructure is energetically more favorable. Although water-splitting remains endothermic, the energy required for dissociation is reduced to 0.338 eV. This result suggests that the the thermodynamic process of water-splitting can be effecttively modulated through doping stratey, facilitating water-splitting and thereby enabling the subsequent HER and OER processes to proceed more efficiently.

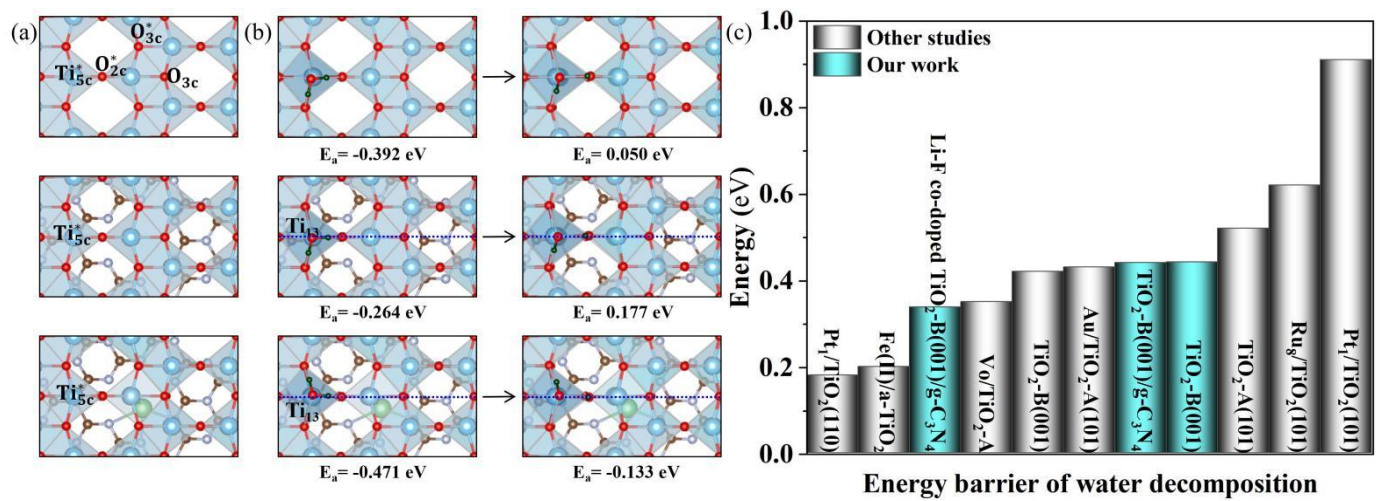

Fig. 1. (a) Surface structural features of TiO$_2$-B(001), TiO$_2$-B(001)/g-C$_3$N$_4$ and Li-F co-doped TiO$_2$-B(001)/g-C$_3$N$_4$. (b) The most stable configurations for molecule adsorption and dissociation adsorption of water on various TiO$_2$-B(001) surfaces. Adsorption energies are also given below each geometry. (c) Comparative energy profiles for water-splitting on TiO$_2$-based materials, referenced to previous studies.[22-29]

Briefly, the water-splitting on $TiO_2$-B(001) surface, $TiO_2$-B(001)/g-$C_3N_4$, and Li-F co-doped $TiO_2$-B(001)/g-$C_3N_4$ heterostructures is found to be an endothermic process. Water-splitting is not kinetically controlled, but rather thermodynamically driven. The energy required for water-splitting on these surfaces are 0.442 eV, 0.441 eV, and 0.338 eV, respectively, showing a gradual decreasing trend. This result suggests that the formation of the $TiO_2$-B(001)/g-$C_3N_4$ heterostructure and Li-F co-doping significantly promote water-splitting, thereby facilitating the subsequent HER and OER processes.

To validate our proposed design strategy for constructing and modulating heterostructures to enhance photocatalytic water-splitting, we further analyzed the energy required for water-splitting in previous studies on $TiO_2$-based nanomaterials, as illustrated in Fig. 1(c). A comparison of the energy requirements for water-splitting on various $TiO_2$-based materials reveals that the energy needed for water-splitting on $TiO_2$-B(001) surface and the pristine heterostructure falls within a moderate range.[22-29] However, the energy required for water-splitting on the proposed Li-F co-doped heterostructure is significantly reduced. Overall, the heterostructure and the Li-F co-doped heterostructure designed and proposed in the current study demonstrate promising potential as an efficient catalyst for photoelectrochemical water-splitting.

**Proton adsorption at different surfaces and interfaces**

After water-splitting, the adsorbed hydroxyl (*OH) and proton (*H) participate in OER on $TiO_2$-B(001) surface and HER on g-$C_3N_4$ surface, respectively. In a mixture suspension of $TiO_2$-B and g-$C_3N_4$ used as a photocatalyst for water-splitting, protons are adsorbed on the surface of $TiO_2$-B(001) after water-splitting. For a complete hydrogen evolution cycle, the adsorbed protons must migrate from the $TiO_2$-B(001) surface to the g-$C_3N_4$ surface through the interface to facilitate the subsequent hydrogen evolution process. Therefore, the adsorption of proton on various $TiO_2$-B(001) surfaces, their diffusion and migration at the interface of heterostructures, and finally evolution of $H_2$ on various g-$C_3N_4$ surfaces, were systematically investigated. First, some possible adsorption sites of proton on various $TiO_2$-B(001) surfaces were analyzed. Figs. 2(a)~(c) show some possible adsorption sites of proton on the surface of $TiO_2$-B(001), $TiO_2$-B(001)/g-$C_3N_4$ and Li-F co-doped $TiO_2$-B(001)/g-$C_3N_4$ heterostructures. As shown in Fig. 2(a), the presence of dangling bonds at unsaturated $O_{2c}^*$ and $O_{3c}^*$ sites allows proton to easily form polar covalent bonds with them, making these sites the preferred locations for proton adsorption. Therefore, there are two possible adsorption sites on $TiO_2$-B(001) surface. Similarly, in $TiO_2$-B(001)/g-$C_3N_4$ and Li-F co-doped $TiO_2$-B(001)/g-$C_3N_4$ heterostructures, although the symmetry is broken, the primary adsorption sites remain the unsaturated $O_{2c}^*$ and $O_{3c}^*$ sites. Thus, there are eight possible adsorption configurations in total.

Next, some possible adsorption sites of protons on various g-$C_3N_4$ surfaces were also analyzed. As shown in Figs. 2(d)~(f), some possible adsorption sites of protons on surfaces of g-$C_3N_4$, g-$C_3N_4$/$TiO_2$-B(001), and Li-F co-doped g-$C_3N_4$/$TiO_2$-B(001) are presented. As one can see from Fig. 2(d), g-$C_3N_4$ monolayer contains two non-equivalent carbon atoms and three non-equivalent nitrogen atoms, labeled as $C_1$, $C_2$, and $N_1$, $N_2$, and $N_3$, respectively. This means that there are five possible adsorption sites for protons on g-$C_3N_4$ surface. Our previous study showed that nitrogen exhibits higher electronegativity than carbon, and the pyridine-type nitrogen atom ($N_2$, located at the edges of the hole in g-$C_3N_4$) has lone pair electrons, making them effective proton acceptor.[5] As shown in Fig. 2(e), in the g-$C_3N_4$/$TiO_2$-B(001) heterostructure, interfacial interaction causes the g-$C_3N_4$ layer to distort, rendering the pyridine-type nitrogen atoms at different positions non-equivalent. However, the geometry structure retains mirror symmetry (indicated by the dashed lines in the plot), leading to six possible proton adsorption sites on the surface of g-$C_3N_4$/$TiO_2$-B(001) heterostructure. As shown in Fig. 2(f), for Li-F co-doped g-$C_3N_4$/$TiO_2$-B(001) heterostructure, the Li atom occupies a hole within the g-$C_3N_4$ layer. From the perspective of steric hindrance and electron distribution, proton is more likely to attach to pyridine-type nitrogen atom at the edges of the hole in g-$C_3N_4$ layer. For comparison, the possibility of proton adsorption on pyridine-type nitrogen within the hole occupied by the Li dopant is also considered. Consequently, there are also eight possible adsorption sites for proton in this configuration.

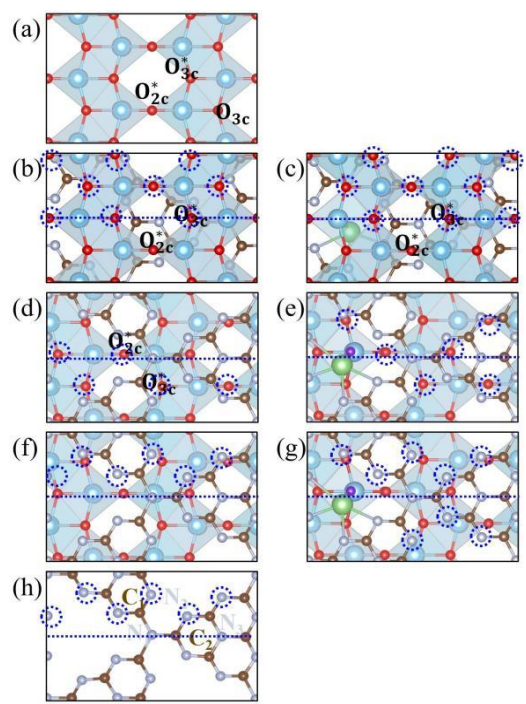

Fig. 2. Possible adsorption sites for proton on different surfaces and interfaces. (a) $TiO_2$-B(001) surface, (b) $TiO_2$-B(001)/g-$C_3N_4$ surface, (c) Li-F co-doped $TiO_2$-B(001)/g-$C_3N_4$ surface, (d) g-$C_3N_4$ surface, (e) g-$C_3N_4$/$TiO_2$-B(001) surface, (f) Li-F co-doped g-$C_3N_4$/$TiO_2$-B(001) surface, (g) $TiO_2$-B(001)/g-$C_3N_4$ interface, (h) Li-F co-doped g-$C_3N_4$/$TiO_2$-B(001) interface.

Finally, the possible adsorption of proton at the interface of the g-$C_3N_4$/$TiO_2$-B(001) and Li-F co-doped g-$C_3N_4$/$TiO_2$-B(001) heterostructures were also investigated, which are shown in Figs. 2(g)~(h). Among them, the possible adsorption sites still are the unsaturated $O^*_{2c}$ and $O^*_{3c}$ sites, but different from the adsorption of proton on various $TiO_2$-B(001) surfaces, proton at the interface tends to orient perpendicularly to enhance the interaction between $TiO_2$-B(001) and g-$C_3N_4$, thereby lowering the total energy of the system. Therefore, the $O^*_{2c}$ and $O^*_{3c}$ sites can also serve as the preferred adsorption sites for proton.

Based on the analysis mentioned above, some possible adsorption configurations were systematically constructed and relaxed, and the adsorption energies of proton at different surfaces or interfaces were also calculated. The adsorption energy is defined as the difference between the total energy of proton adsorption system and the sum of the total energy of the system without proton adsorption and the binding energy of hydrogen molecule. The calculated results show that the adsorption behavior of proton is consistent with our prediction. As shown in Fig. S1, for the adsorption of proton on $TiO_2$-B(001)/g-$C_3N_4$ surface, the adsorption energy at $O^*_{2c}$ site is at least ~0.33 eV lower than that at $O^*_{3c}$ site. The proton and oxygen atom form a polar covalent bond, which is parallel to the interface, with a bond length of 0.98 Å. Therefore, the adsorption configuration of the proton at $O^*_{2c}$ site ($O_{29}$) was selected for the following kinetic analysis. The adsorption configurations of proton at the $TiO_2$-B(001)/g-$C_3N_4$ interface are shown in Fig. S2. It can be seen from the plot that the adsorption energy at $O^*_{3c}$ site (i.e., $O_4$, $O_8$ and $O_{12}$) is at least ~0.01 eV lower than that at $O^*_{2c}$ site (i.e., $O_5$, $O_9$ and $O_2$), and at least ~0.14 eV lower than that at $O_{3c}$ site (i.e., $O_{11}$, $O_3$ and $O_7$). Notably, the polar covalent bond formed between the proton and oxygen atom is perpendicular to the interface, with the bond length extending from 0.98 Å to 1.02 Å, thereby weakening the covalent bond strength. Although the polar covalent bond is weakened, the proton simultaneously forms hydrogen bond (H-bond) with pyridine-type nitrogen on the surface of g-$C_3N_4$ (O-H...N), enhancing the interface interaction. This explains why the adsorption energy at the $O^*_{3c}$ site is the lowest. Consequently, the adsorption configurations at $O^*_{3c}$ site (i.e., $O_4$, $O_8$ and $O_{12}$) were chosen for the following kinetic analysis. However, it is worth noting that the H-bond length in the $O_{12}$ configuration is 0.03 Å shorter than that in the $O_4$ and $O_8$ configurations, making the $O_{12}$ configuration the preferred choice for further kinetic analysis.

The possible adsorption configurations of protons on the surface and interface of the Li-F co-doped $TiO_2$-B(001)/g-$C_3N_4$ heterostructure are plotted in Figs. S3(a)~(b). As one can see from Fig. S3(a), the adsorption of proton at $O^*_{2c}$ site (i.e., $O_{29}$ and $O_{22}$) remain dominant, with the adsorption energy being ~0.33 eV lower than that at other sites. Similar to the un-doping system, the proton forms a polar covalent bond with the oxygen atom, with a bond length of 0.98 Å, which is also parallel to the interface. Therefore, the adsorption configuration of proton at $O^*_{2c}$ site ($O_{29}$) was selected for the following kinetic analysis. It can be seen from Fig. S3(b) that the adsorption energy of proton at $O^*_{2c}$ site ($O_9$) is lower than that at $O^*_{3c}$ site ($O_{12}$) and significantly lower than that at $O_{3c}$ site ($O_6$). The adsorption of proton at the interface remains perpendicular to enhance the interface interaction. It is worth pointing out that the proton not only forms a polar covalent bond with the oxygen atom, but also establishes H-bond with the interfacial dopant F and nitrogen atom on the surface of g-$C_3N_4$. The H-bond lengths in the configurations of $O_9$, $O_{12}$ and $O_6$ are 1.88 Å, 1.75 Å and 1.76 Å, respectively. From the geometric perspective, the adsorption configuration of proton at $O^*_{3c}$ site ($O_{12}$) is more favorable for interfacial migration of proton. Therefore, both $O_9$ and $O_{12}$ configurations were used for subsequent kinetic migration analysis.

After proton migration at the interface, the HER process is finally achieved on the surface of g-$C_3N_4$. Therefore, the adsorption configurations and thermodynamic properties of proton on different g-$C_3N_4$ surfaces have also been systematically investigated. The various proton adsorption isomers are shown in Fig. S4. For comparison, the adsorption of proton on free-standing g-$C_3N_4$ surface was also systematically examined. The calculation results showed that pyridine-type nitrogen serves as an excellent proton acceptor, with the lowest adsorption energy observed at these sites, as shown in Fig. S4(a). For proton adsorption on the surface of g-$C_3N_4$/$TiO_2$-B(001) heterostructure, the presence of mirror symmetry (indicated by blue dash line), results in equal adsorption energies for many equivalent sites, such as $N_{16}$ and $N_8$, $N_{11}$ and $N_{14}$, $N_2$ and $N_{12}$, $N_3$ and $N_{15}$, etc., as depicted in Fig. S4(b). Overall, pyridine-type nitrogen remains the preferredproton acceptor. Therefore, the adsorption of proton at $N_{16}$ site was selected for free energy analysis in subsequent kinetic migration and HER process. In the Li-F co-doped heterostructure, the adsorption energy of pyridine-type nitrogen $N_2$ site, located at the edge of hole occupied by Li-F co-dopants, is 0.26 eV higher than that of the non-occupied pyridine-type nitrogen $N_{16}$ site, as we can see from Fig. S4(c). Therefore, the subsequent kinetic migration process and HER process primarily occur at $N_{16}$ site.

Briefly, proton is more easily adsorbed on $O^*_{2c}$ site to form polar covalent bond which is parallel to the interface of different $TiO_2$-B(001) surfaces. Similarly, proton tends to adsorb on pyridine-type nitrogen atom to form polar covalent bond which is also parallel to the interface of different g-$C_3N_4$ surfaces. At the interfaces of heterostructures, interfacial interaction causes proton to preferentially adsorb at $O^*_{3c}$ site in g-$C_3N_4$/$TiO_2$-B(001) heterostructure or $O^*_{2c}$ site in Li-F co-doped g-$C_3N_4$/$TiO_2$-B(001) heterostructure, forming polar covalent bond which is perpendicular to the interface. Additionally, proton is more likely to form H-bond (O-H...N), further stabilizing the system. In order to comprehensively examine the migration process and energy barrier for proton comprehensively, the most stable adsorption configurations of proton on various surfaces or interfaces are shown in Fig. 3(a), with their corresponding adsorption energies plotted in Fig. 3(b). It can be observed from the plots that the proton is more easily adsorbed on the surface of g-$C_3N_4$ (-0.495 eV) than on the surface of $TiO_2$-B(001) (-0.089 eV), indicating that g-$C_3N_4$ is an effective proton acceptor. The adsorption energies of proton on pristine $TiO_2$-B(001) surface, $TiO_2$-B(001)/g-$C_3N_4$ surface and Li-F co-doped $TiO_2$-B(001)/g-$C_3N_4$ surface are -0.089eV, 0.041eV and -0.418 eV, respectively. This result shows that Li-F co-doping enhances the adsorption strength of proton on the surface of heterostructures, ensuring stable proton adsorption after water-splitting. Furthermore, the adsorption energies of proton on the surface of $TiO_2$-B(001)/g-$C_3N_4$ heterostructure, interface and g-$C_3N_4$ surface are 0.041eV, -0.380eV and -0.586 eV, respectively. The adsorption energy gradually decreases from the $TiO_2$-B(001) surface to the interface and then to the g-$C_3N_4$ surface. A similar trend is observed in the Li-F co-doped $TiO_2$-B(001)/g-$C_3N_4$ heterostructure, with the adsorption energy at each site being lower than that in the pristine heterostructure, as shown in Fig. 3(b). This indicates that, after water-splitting, proton tends to migrate from $TiO_2$-B(001) surface to the interface, and then to g-$C_3N_4$ surface, particularly in the Li-F co-doped system. Finally, the adsorption energies of proton on g-$C_3N_4$, g-$C_3N_4$/$TiO_2$-B(001) and Li-F co-doped g-$C_3N_4$/$TiO_2$-B(001) surfaces are -0.418eV, -0.586eV and -0.646 eV, respectively, also exhibiting a gradually decreasing trend.

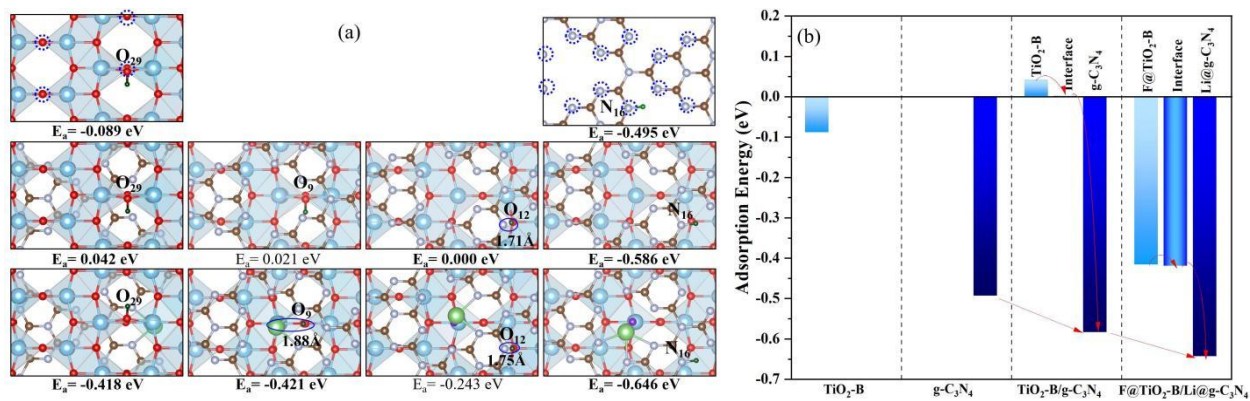

**Fig. 3.** (a) The most stable adsorption configurations of proton on different surfaces and interfaces. (b) Adsorption energies for proton on different surfaces and interfaces.

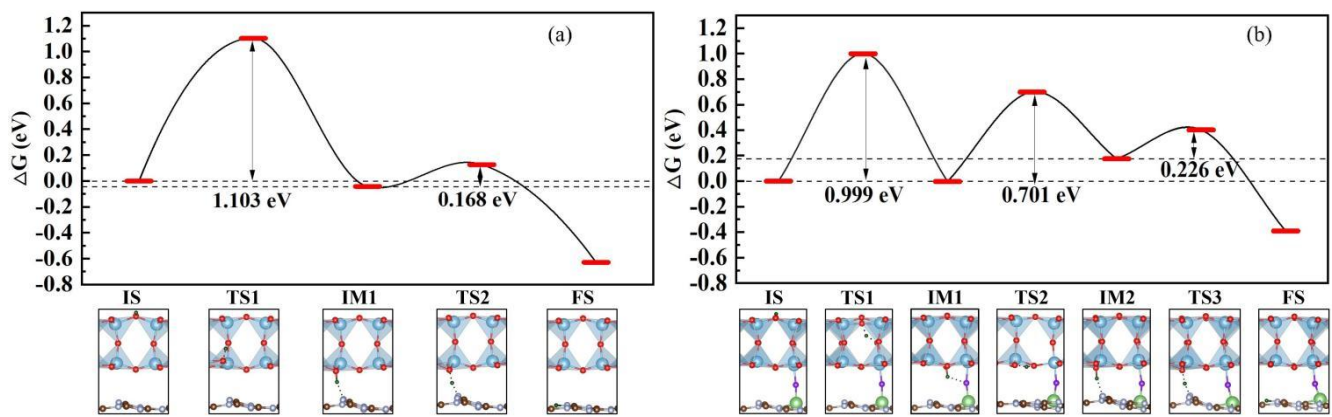

**Fig. 4.** Proton migration reaction pathways, energy barriers and geometries. (a) From $TiO_2$-B(001) surface to interface and then to g-$C_3N_4$ surface. (b) From $TiO_2$-B(001) surface to Li-F co-doped interface and then to g-$C_3N_4$ surface.

**Proton migration from $TiO_2$-B(001) to g-$C_3N_4$ surface**

Based on the thermodynamic stability analysis discussed above, it is found that in the heterostructures, proton tends to migrate from $TiO_2$-B(001) surface to the interface, and then to g-$C_3N_4$ surface. To verify this trend, proton migration pathways and migration energy barriers were calculated by using the CI-NEB method. According to the thermodynamic analysis, the proton migration path in g-$C_3N_4$/$TiO_2$-B(001) heterostructure follows the route $O_{29}\rightarrow O_{12}\rightarrow N_{16}$, involving two transition states: the transition state from the $TiO_2$-B(001) surface to the interface (TS1) and the transition state from the interface to g-$C_3N_4$ (TS2). In Li-F co-doped g-$C_3N_4$/$TiO_2$-B(001) heterostructure, the proton migration path is $O_{29}\rightarrow O_9 \rightarrow O_{12}\rightarrow N_{16}$, including the transition state from $TiO_2$-B(001) surface to interface (TS1), the transition state of the proton interfacial diffusion (TS2) and the transition state from interface to g-$C_3N_4$ surface (TS3). The corresponding proton migration pathways and energy barriers are plotted in Fig. 4.

From Fig. 4(a), it can be observed that the migration of proton from $TiO_2$-B(001) surface to the interface requires overcoming an energy barrier of 1.103 eV in $TiO_2$-B(001)/g-$C_3N_4$ heterostructure. The TS1 corresponds to cleavage (where O acts as a proton donor) and recombination (where O acts as a proton acceptor) of the polar O-H covalent bond. The energy barrier for proton migration from interface to g-$C_3N_4$ surface is only 0.168 eV, with TS2 corresponding to the cleavage of the polar O-H bond (O acts as proton donor) and formation of the polar N-H bond (N acts as proton acceptor).

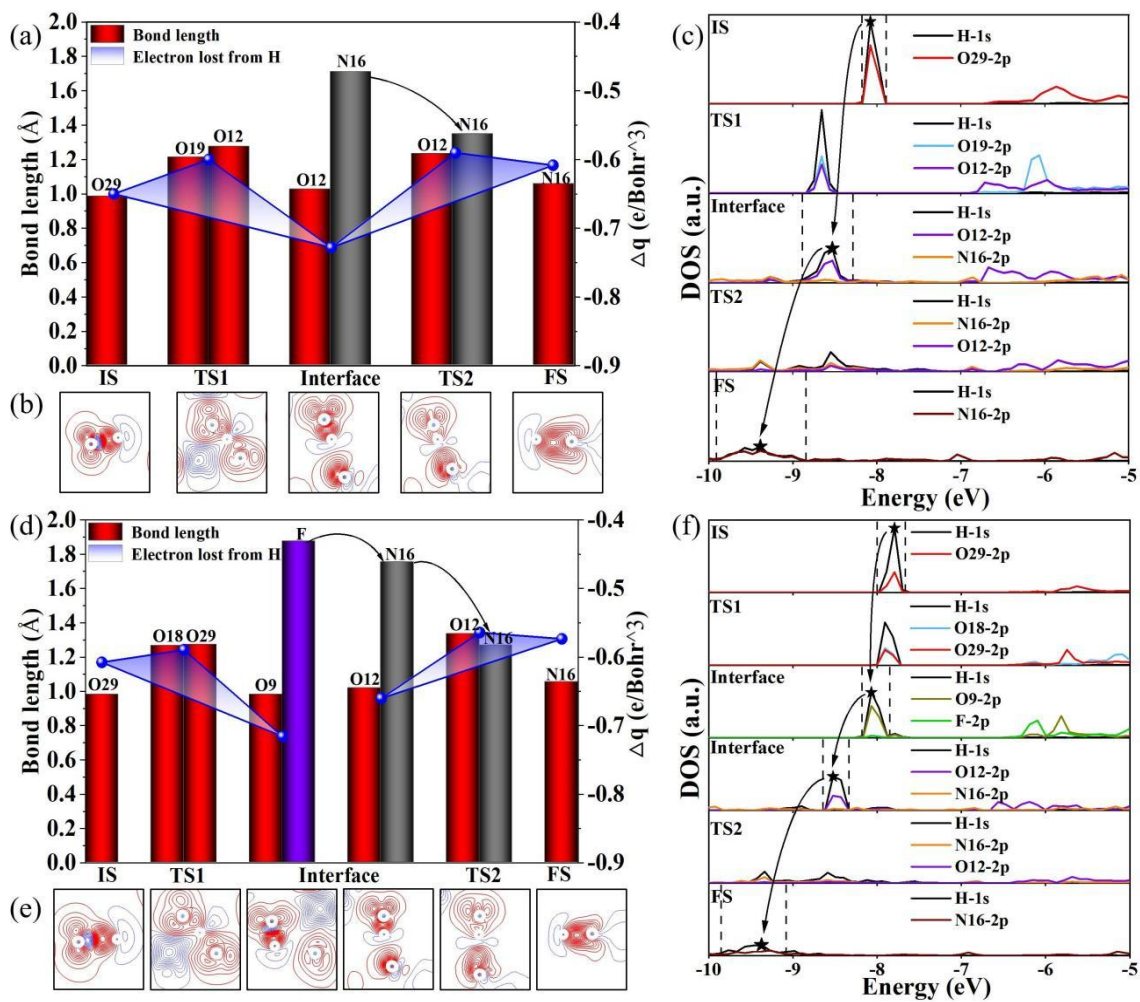

**Fig. 5.** (a) Electron lost from proton, bond lengths and (b) two-dimensional (2D) charge density difference cutting through H-O or H-N bonds with range of ±0.007 e/Å³ in g-C₃N₄/TiO₂-B(001) heterostructure. (c) The partial density of states for proton and its adjacent atoms in g-C₃N₄/TiO₂-B(001) heterostructure during proton migration. (d) Electron lost from proton, bond lengths and (e) two-dimensional (2D) charge density difference cutting through H-O or H-N bonds with range of ±0.007 e/Å³ in Li-F co-doped g-C₃N₄/TiO₂-B(001) heterostructure. (f) The partial density of states for proton and its adjacent atoms in Li-F co-doped g-C₃N₄/TiO₂-B(001) heterostructure during proton migration.

In the Li-F co-doped g-C₃N₄/TiO₂-B(001) heterostructure, the migration of proton from TiO₂-B(001) surface to the interface needs to overcome an energy barrier of 0.999 eV, and the TS1 corresponds to the cleavage (O as proton donor) and recombination (O as proton acceptor) of the polar O-H covalent bond. Different from the pristine heterostructure, proton diffusion occurs at the interface of the Li-F co-doped g-C₃N₄/TiO₂-B(001) heterostructure. The adsorption energy of proton on the most stable $O^*_{2c}$ site ($O_9$) is 0.178 eV lower than that on the metastable $O^*_{3c}$ site ($O_{12}$). The diffusion of the proton from the most stable site to the metastable site involves an energy barrier of 0.701 eV, as one can see from Fig. 4(b). The highest barrier point resembles the proton diffusion from TiO₂-B(001) surface to the interface, requiring energy to break the polar covalent bond. Thus, TS2 also corresponds to cleavage and recombination of the polar O-H covalent bond, where O acts as either a proton donor or acceptor. It should be noted that during the diffusion from $O_9$ to $O_{12}$, the system first undergoes an energy decrease of 0.14 eV, reaching an intermediate state, then rises to the transition state (TS), and finally decreases to the final state, which one can see from Fig. S5. Then, proton from $O^*_{3c}$ site migrates to pyridine-type nitrogen site ($N_{16}$) on the surface of g-C₃N₄ via a H-bond (O-H···N), with an associated energy barrier of 0.226 eV. TS3 corresponds to cleavage of the polar O-H bond (O as proton donor) and the formation of the polar N-H bond (N as proton acceptor). Therefore, the energy barrier for proton migration from the interface to g-C₃N₄ surface increases to 0.226 eV.

In short, the energy barrier for proton migration from the TiO₂-B(001) surface to interface is the determining step influencing HER performance in pristine g-C₃N₄/TiO₂-B(001) and Li-F co-doped g-C₃N₄/TiO₂-B(001) heterostructures. However, the diffusion energy barrier of approximately 1 eV is not sufficiently high to inhibit the HER process. With precise experimental controls, such as heating and stirring, hydrogen evolution on the surface of the g-C$_3$N$_4$/TiO$_2$-B(001) heterostructures can be readily achieved in experiment.

To understand the characteristics of the energy barrier during proton migration or diffusion, detailed analyses of bond length, Bader charge, charge density differences and partial density of state (PDOS) were conducted for the initial state (IS), transition state (TS), interface configuration, and final state (FS), as illustrated in Fig. 5. It can be observed from Figs. 5(a)~(b) that in the IS structure, the bond length between proton and oxygen is 0.98 Å, with the proton losing 0.65|e|. The two-dimensional (2D) charge density difference distribution shows significant charge accumulation between the proton and the oxygen atom, and the PDOS shows strong hybridization between the proton and oxygen around -8.0 eV in g-C$_3$N$_4$/TiO$_2$-B(001) heterostructure. These findings indicate the formation of a strong polar covalent bond between the proton and the oxygen atom.

In TS1 structure, the distances between the proton and its adjacent oxygen atoms increase to 1.21 Å and 1.27 Å, respectively, and the electron loss of the proton decreases to 0.60|e|. The 2D charge density contours become sparser compared to the IS structure, and the interaction between the proton and oxygen weakens. The PDOS also shows reduced hybridization between the proton and its neighboring oxygen atoms. These results indicate that the interaction between proton and the heterostructure system is significantly weakened, resulting in a substantial increase in the energy of TS1, and consequently, a higher energy barrier for proton migration. In other words, the strong polar covalent bond is the primary factor contributing to hte high energy barrier for proton migrating from TiO$_2$-B(001) surface to the interface.

At the interface, in addition to forming a strong polar covalent bond (1.03 Å) with the oxygen atom, the proton also forms a H-bond with pyridine-type nitrogen atom in g-C$_3$N$_4$. The H-bond length is 1.71 Å and the O-H…N bond angle is 165.25°, approaching linear arrangement. This geometry is highly favorable for proton migration from the interface to g-C$_3$N$_4$ surface. Based on the denser 2D charge density contours and stronger hybridization in the PDOS within the energy range of -9.0~-8.5 eV, it can be inferred that the proton forms a moderate-strength H-bond with pyridine-type nitrogen, stabilizing hte system. The charge loss from the proton recovers to 0.73|e| at this stage.

In the TS2 structure, the O…H bond length is 1.23 Å, the N…H bond length is 1.35 Å, and the O…H…N bond angle increases to 166.90°. The 2D charge density contours are sparse, and the PDOS shows weak hybridization between the proton and oxygen, as well as between the proton and the pyridine-type nitrogen atom, within the energy range of -9.0~-8.5 eV. This indicates the formation of weak chemical bonds, and electron loss from the proton decrease further to 0.59|e|. In short, the presence of the H-bond results in a low migration barrier for protons moving from the interface to g-C$_3$N$_4$ surface.

Finally, in the FS structure, the proton is adsorbed on the pyridine-type nitrogen in g-C$_3$N$_4$ with an N-H bond length of 1.02 Å. The 2D charge density contours reveal charge accumulation only between the pyridine-type nitrogen and the proton, implying that the formation of a polar N-H covalent bond. In addition, as one can see from Fig. 5(b), the PDOS of the proton is more delocalized, with strong hybridization between the proton and the pyridine-type nitrogen in the energy range of -10.0 eV~-9.0 eV. At this stage, the system is most stable, and correspondingly, the proton loses the least amount of charge (0.61|e|) due to the lower electronegativity of nitrogen compared to oxygen. This provides favorable conditions for the subsequent HER reaction.

Figs. 5(c)~(d) systematically illustrate the proton migration pathways and corresponding energy barriers within the Li-F co-doped TiO$_2$-B(001)/g-C$_3$N$_4$ heterostructure, accompanied by comprehensive analyses of bond length, Bader charge, charge density difference and projected state density analysis for the initial state, transition state, interface configuration and final state. The fundamental mechanism governing proton migration, energy barrier changes, and their physical origin exhibit remarkable parallels with of those observed in pristine TiO$_2$-B(001)/g-C$_3$N$_4$ heterostructure. For example, it can be seen from Fig. 5(c), the pronounced polarization of the O-H covalent bond leads to a substantial migration barrier (0.999 eV) for proton transfer from the TiO$_2$-B(001) surface to the interface, while the weakened H-bond interaction reduces the migration barrier (0.226 eV) for proton migration from the interface to g-C$_3$N$_4$ surface. However, in the Li-F co-doped TiO$_2$-B(001)/g-C$_3$N$_4$ heterostructure, proton can not migrate directly from the interface to g-C$_3$N$_4$ surface after transferring from TiO$_2$-B(001) surface. Instead, the migration process is facilitated by interfacial proton diffusion. Specifically, as mentioned above, this diffusion involves the movement of proton from O$_9$ site to O$_{12}$ site, accompanied by three key steps: (1) the cleavage of the H-bond (O-H…F), (2) the reorientation of the proton (i.e., the polar covalent O-H bond shifts from a perpendicular to a parallel orientation relative to the interface), and (3) the cleavage and reformation of the polar O-H covalent bond. These steps collectively require the proton to overcome an energy barrier of 0.701 eV to complete the diffusion process. It should be noted, however, that the overall proton migration process is not solely governed by interfacial diffusion. In addition, it can be seen from Fig. 5(d) that the hybridization between the proton and the oxygen atom on the TiO$_2$-B(001) surface, the interfacial oxygenatom, and the nitrogen atom on the g-C$_3$N$_4$ surface gradually shifts to lower energy. This shift indicates a progressive increase in the stability of proton binding with surface oxygen atom, interfacial oxygen atom, and surface nitrogen atom. This trend is consistent with the trend in adsorption energy changes, providing further evidence for the feasibility of proton migration within heterostructures.

Comparing Figs. 4(a)~(b), it can be observed that the energy barrier for proton migration from the TiO$_2$-B(001) surface to the interface in the Li-F co-doped heterostructure is reduced by ~0.1 eV compared to the pristine heterostructure. Carefully observing Fig. 5(a) and Fig. 5(c), it can be seen that in the Li-F co-doped heterostructure, proton on the surface of the TiO$_2$-B(001) loses 0.61|e|, while in the pristine heterostructure, the proton loses 0.65|e|. This result indicates that the polar O-H covalent bond is weaker in the Li-F co-doped heterostructure

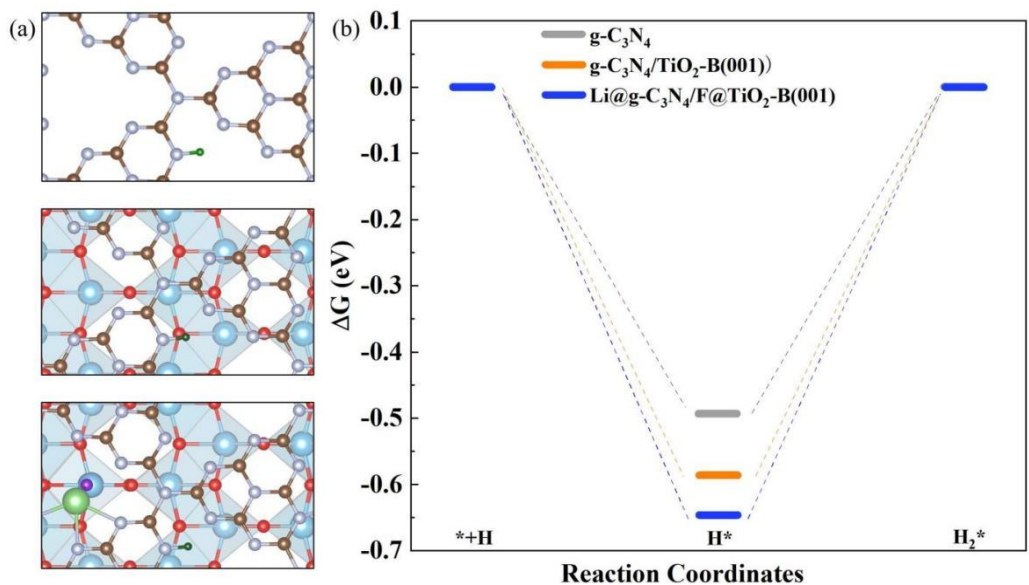

**Fig. 6.** The most stable adsorption configurations of proton on g-$C_3N_4$, g-$C_3N_4$/$TiO_2$-B(001) and Li-F co-doped g-$C_3N_4$/$TiO_2$-B(001) surfaces. (b) HER free energy diagrams for proton on g-$C_3N_4$, g-$C_3N_4$/$TiO_2$-B(001) and Li-F co-doped g-$C_3N_4$/$TiO_2$-B(001) surfaces.

than the polar O-H covalent bond in pristine heterostructure. The weakening for the polar O-H covalent bond is further supported by PDOS analysis, as shown in Fig. 5(b) and Fig. 5(d), which demonstrates a significant reduction in hybridization between the proton and oxygen in the Li-F co-doped heterostructure. Consequently, the energy required to cleave the polar O-H covalent bond is correspondingly lower. These results indicate that the reduced energy barrier for proton migration from $TiO_2$-B(001) surface to the interface in the Li-F co-doped heterostructure is primarily attributed to the decreased strength of the polar O-H covalent bond.

On the other hand, according to previous studies,[13] the interfacial polarization (built-in electric field) in the Li-F co-doped heterostructure is stronger than that in pristine heterostructure. This enhanced interfacial polarization provides an additional driving force for proton migration, facilitating the movement of proton from the $TiO_2$-B(001) surface to the interface and thereby reducing the migration energy barrier. It is found that the energy barrier for proton migration from the interface to g-$C_3N_4$ surface in the Li-F co-doped heterostructure (0.226 eV) is higher than that in the pristine heterostructure (0.168 eV). In other words, Li-F co-doping increases the energy barrier for proton migration from the interface to the g-$C_3N_4$ surface. This is mainly due to the interfacial polarization directed from g-$C_3N_4$ to $TiO_2$-B(001) in the Li-F co-doped heterostructure, which partially inhibits proton migration. Notably, the polar O-H covalent bond length (1.02 Å), the hydrogen bond length of O-H...N (1.75 Å), and the electrons loss from the proton (0.66|e|) in the Li-F co-doped heterostructure are similar to those in the pristine heterostructure (i.e., the polar O-H covalent bond length (1.03 Å), the hydrogen bond length of O-H...N length (1.71 Å), and electrons loss from the proton (0.73|e|)). Despite this, the Li-F co-doped heterostructure is more favorable for proton migration and the subsequent HER process compared to the pristine heterostructure.

**HER process on different g-$C_3N_4$ surfaces**

Based on the thermodynamic and kinetic analyses discussed in the previous section, it can be concluded that the $TiO_2$-B(001)/g-$C_3N_4$ heterostructure and the Li-F co-doped $TiO_2$-B(001)/g-$C_3N_4$ heterostructure are promising photocatalysts for hydrogen evolution. In order to evaluate their HER performance, the Gibbs free energy of hydrogen adsorption ($\Delta G(*H)$) was calculated according to the Volmer-Heyrovsky mechanism. The Gibbs free energy of hydrogen adsorption ($\Delta G(*H)$) is widely regarded as a reliable descriptor and is frequently used to describe the intrinsic activity of photocatalysts to HER. According to the Sabatier principle,[30] the optimal thermoneutral active site, where $\Delta G(*H)$ is close to zero, promotes both the adsorption and desorption process of HER. In heterostructured hydrogen evolution photocatalysts, the HER process involves strong proton adsorption on the catalyst surface (e.g., $\Delta G(*H) < 0$) and efficient desorption of $H_2$ (e.g., $\Delta G(*H_2) > 0$).

Based on previous thermodynamics analyses, proton tends to adsorb on pyridinic-type nitrogen on the g-$C_3N_4$ surface. The adsorption configurations of proton on different g-$C_3N_4$ surfaces are plotted in Fig. 6(a). To further evaluate the HER performance of these three photocatalysts, the Gibbs free energy diagram for HER under conditions of U = 0 V and pH = 0 is plotted in Fig. 6(b). From the free energy diagram, it can be observed that $\Delta G(*H)$ is -0.495 eV for proton on the pure g-$C_3N_4$ surface, indicating stable proton adsorption. $\Delta G(*H)$ decreases to -0.586 eV for proton on the g-$C_3N_4$/$TiO_2$-B(001) heterostructure, suggesting that the charge transfer at the heterostructure interface enhances proton adsorption but slightly hinders desorption. In the Li-F co-doped g-$C_3N_4$/$TiO_2$-B(001) heterostructure, $\Delta G(*H)$ further decreases to -0.646 eV, implying that the localized interfacial electric field introduced by Li-F doping optimizes the electronic distribution, making proton adsorption more stable. However, excessively strong adsorption can inhibit $H_2$ production. From the above analysis, it is clear to see that the g-$C_3N_4$/$TiO_2$-B(001) heterostructure exhibits more favorable proton adsorption and HER activity, while Li-F doping may further enhance proton adsorption, but could impede $H_2$ desorption. Therefore, appropriate interfacial design and a balanced doping strategy are crucial for optimizing HER performance.

Considering the thermodynamic and kinetic processes of proton migration, although Li-F co-doping increases the energy barrier for proton migration from the interface to the g-$C_3N_4$ surface, this process is not the determining process. The key factor influencing HER performance is the migration of proton from the $TiO_2$-B(001) surface to the interface. Li-F co-doping reduces the energy barrier for proton migration, making it more favorbale for proton to migrate from the $TiO_2$-B(001) surface to the interface. Once proton migrates to the g-$C_3N_4$ surface, it participates in the subsequent HER process, where the critical step affecting HER performance is the desorption of $H_2$. This is because the desorption energy barrier of $H_2$ from pristine and Li-F co-doped heterostructures is higher than that from the g-$C_3N_4$ surface. Nevertheless, the desorption barrier of $H_2$ remains acceptable compared to the proton migration barrier. The results of the current study demonstrate that the g-$C_3N_4$/$TiO_2$-B(001) heterostructure and the Li-F co-doped g-$C_3N_4$/$TiO_2$-B(001) hterostructure are highly effective photocatalytic hydrogen evolution catalysts.

## Conclusions

Based on density functional theory, this study systematically investigates the water-splitting, proton migration behavior, and hydrogen evolution reaction performance of g-$C_3N_4$/$TiO_2$-B(001) and Li-F co-doped g-$C_3N_4$/$TiO_2$-B(001) heterostructures. The results show that water molecules can be stably adsorbed on the surfaces of both g-$C_3N_4$/$TiO_2$-B(001) and Li-F co-doped heterostructures. The energies required for water-splitting on surfaces of $TiO_2$-B(001), g-$C_3N_4$/$TiO_2$-B(001) heterostructure, and Li-F co-doped heterostructure are 0.442 eV, 0.441 eV, and 0.338 eV, respectively, indicating that both heterostructure formation and Li-F co-doping synergistically promote water-splitting.

Proton migration from the $TiO_2$-B(001) surface to the g-$C_3N_4$ surface exhibits a significant decreasing trend. Proton migration from $TiO_2$-B(001) to the interface involves a high energy barrier of 1.103 eV, primarily due to the breaking and reformation of the polar H-O covalent bonds. In contrast, migration from the interface to the g-$C_3N_4$ surface is facilitated by interfacial O…H…N hydrogen bonding, reducing the energy barrier to 0.168 eV. Furthermore, the Li-F co-doping weakens the O-H bond strength and enhances interfacial polarization, lowering the energy barrier for proton migration from $TiO_2$-B(001) to the interface to 0.999 eV. Additionally, proton diffusion at the interface introduces an energy barrier of 0.701 eV. Although this interface diffusion barrier is relatively high, the overall energy barrier remains low.

Finally, the HER free energy diagrams reveal that the g-$C_3N_4$/$TiO_2$-B(001) heterostructure achieves a balance between *H adsorption and desorption, with an adsorption free energy $\Delta G(*H)$ of -0.586 eV. Li-F co-doping further promotes proton adsorption, reducing $\Delta G(*H)$ to -0.646 eV. In summary, the g-$C_3N_4$/$TiO_2$-B(001) and Li-F co-doped heterostructure exhibit excellent HER catalytic activity due to their low energy for water-splitting, low proton migration energy barrier, and optimal *H adsorption and desorption characteristics. This study elucidates the atomic-scale mechanism of proton migration at the heterostructure interface, providing valuable theoretical insights for the rational design of efficient HER catalysts.

## Author contributions

Shuhan Tang: methodology, investigation, conceptualization, data curation, writing-original draft. Qi Jiang: writing-review & editing. Shuang Qiu: writing-review & editing. Hanyang Ji: writing-review & editing. Xiaojie Liu: conceptualization, data curation, validation, resources, writing-review & editing, supervision.

## Conflicts of interest

There are no conflicts to declare.

## Data availability

All relevant data are provided within the manuscript and its ESI.†

## Acknowledgements

The authors acknowledge the support by the National Natural Science Foundation of China under Grant No. 11574044, Open Project of Key Laboratory for UV-Emitting Materials and Technology of Ministry of Education (No. 130028608) and the Fundamental Research Funds for the Central Universities. The calculations were also performed on high performance computing platform of Northeast Normal University.

## Notes and references